\documentclass[a4paper,12pt]{article}
\pdfoutput=1
\usepackage{amsmath,amssymb}
\usepackage{amstext}
\usepackage{graphicx}
\usepackage{bm}
\usepackage{epsfig}%
\title{Redundancy of the cosmological evolution equations and its relationship with the initial conditions}
\bigskip
\author{Kaushik Bhattacharya$^\ddagger$, Dipanjan Dey$^\dagger$, Priyanka Saha$^\$$
\thanks{$^\ddagger$kaushikb@iitk.ac.in, $^\dagger$deydipanjan7@gmail.com, $^\$$priyankas21@iitk.ac.in }
\\
\normalsize
$^\ddagger$,$^\$$ Department of Physics, Indian Institute of Technology, Kanpur\\
\normalsize
Uttar Pradesh 208016, India\\
\normalsize
$^\dagger$ Beijing Institute of Mathematical Sciences and Applications\\
\normalsize
Beijing 101408, China
}
\begin{document}
\maketitle
\begin{abstract}

  It is known that in Friedmann-Lemaitre-Robertson-Walker cosmology one has more number of dynamical equations, compared to the number of unknown variables. This fact makes some equations redundant. The situation becomes complicated because all the relevant differential equations in cosmology are not of the same order. In this article we study the fate of the redundant equations. We show that this redundancy
is inevitable in general relativity. It is shown that this redundancy is primarily responsible for a special role of one of the Friedmann equations, which constrains the initial values of the problem. Our method of analyzing the dynamical structure of the theories relies on an operational approach and can be generalized further.

\end{abstract}
\section{Introduction}

While going through the basics of cosmology one first learns to tackle the dynamical equations, commonly called the Friedmann equations in Friedmann-Lemaitre-Robertson-Walker (FLRW) spacetime \footnote{The standard paradigm of introductory cosmology is well addressed in Refs.~\cite{Kolb:1990vq, Mukhanov:2005sc, Dodelson:2003ft} whereas Refs.~\cite{Zee:2013dea, Hartle:2021pel, Weinberg:1972kfs} focusses more on general relativity. Some excellent pedagogical articles on introductory cosmology can also be found in Refs.~\cite{Nemiroff:2007xs, Sonego:2011rb, Faraoni:1999qu, ReFiorentin:2023bhk, Christiansen:2012fy}.}. These Friedmann equations are nonlinear ordinary differential equations which originate from the Einstein equations in general relativity (GR).
While dealing with these equations one first observes that even in the simplest situation, where the universe only contains one barotropic fluid, there is a problem of redundancy. There are simply more number of cosmological evolution equations than required. The situation is more challenging as all the available differential equations are not of the same order. Naturally the question arises why such a problem appears at all and how are we going to obtain proper cosmological development from these equations in a consistent manner. In this article we will primarily be addressing this redundancy problem in cosmology. We will first show that this redundancy problem is intimately related to the initial value problem in cosmology. It turns out that any arbitrary initial condition will not produce proper cosmological dynamics, only initial conditions satisfying a particular constraint will produce proper dynamics. Then we show that the redundancy problem in cosmology has its own logic in GR.       

The interesting point which becomes apparent from the present article is related to the fact that we can learn many important properties related to cosmological development from an operational approach, where our aim is solely to solve the dynamical equations unambiguously. Such an approach ultimately leads to a thorough understanding of the logical structure of the theory.

\section{A single barotropic fluid in FLRW spacetime and the Einstein equation}

In this work we will assume that the FLRW spacetime is seeded by a single component barotropic fluid whose equation of state (EoS): $P=\omega \rho\,,$ where $\omega$,  is a dimensionless quantity in our system of units (where $c=1$). Here $\rho$ is the energy-density of the fluid and $P$ specifies its isotropic pressure. The EoS parameter is a constant. In presence of a barotropic fluid and cosmological constant $\Lambda$, the Einstein equations of FLRW spacetime, with a scale-factor $a(t)$, are given as:
\begin{eqnarray}
\frac{\dot{a}^2}{a^2} + \frac{k}{a^2} &=& \frac{8\pi G}{3}\rho + \frac{\Lambda}{3}\,,
\label{fried1}\\
2\frac{\ddot{a}}{a} + \frac{\dot{a}^2}{a^2} + \frac{k}{a^2} &=& -8\pi G P + \Lambda\,,
\label{fried2}
\end{eqnarray}
where the dot specifies a derivative with respect to time, $t$, and $k$ is a constant which can take values $\pm 1$ or $0$. If $k=0$, the curvature of the spatial hypersurfaces in FLRW spacetime is zero, on the other hand $k= +1,\,-1$, specifies that the spatial hypersurfaces are positively curved (as on the surface of a sphere) or negatively curved (as in a saddle), respectively.

Subtracting Eq.~(\ref{fried1}) from Eq.~(\ref{fried2}) one gets another dynamical equation in FLRW cosmology:
\begin{eqnarray}
\frac{\ddot{a}}{a}= -\frac{4\pi G}{3}(\rho + 3P) + \frac{\Lambda}{3}\,.
\label{fried3}
\end{eqnarray}
In this work we will call Eq.~(\ref{fried1}) as the Friedmann equation, the second equation, Eq.~(\ref{fried2}), will be called as the pressure equation and the last equation will be named as the acceleration equation.

Except the above three equations we have 
the energy-momentum conservation condition in GR, which is an outcome of the Bianchi identity. In FLRW spacetime this condition becomes
\begin{eqnarray}
\dot{\rho} + 3 \frac{\dot{a}}{a}(\rho + P)=0\,.
\label{conteqn}
\end{eqnarray}
We will call Eq.~(\ref{conteqn}) as the continuity equation in this work. As like the continuity equations in other branches of physics, this equation also says how the energy density of the system continuously adjusts itself so that the net energy-momentum of the system is conserved. 

There are now four dynamical equations for the development of the FLRW spacetime.
It must be noted that, although we have two independent differential equations coming out from the Einstein field equation (the Friedmann equation and the the pressure equation), they cannot be simultaneously solved as none of them contain derivatives of $\rho$ or $P$. In such a case one is bound to work with the continuity equation as it is the only one which has a time derivative of $\rho$. There are three functions to be determined: $a(t)$, $\rho(t)$ and $P(t)$. Using the EoS, $P(t)$ can be expressed in terms of the energy density and hence, there are only two unknown functions.

Is it always possible that the four differential equations agree during $t_i\le t \le t_f$ while the actual dynamics is produced  by only two differential equations satisfying some specific  initial conditions, keeping $a(t)>0$ and $\rho(t)>0$ throughout the time interval $[t_i\,,\,t_f]$? The answer turns out to be positive for a certain set of initial conditions. For initial conditions which are outside the preferred set, all the four equations will never match and the system will behave as an overdetermined system. To present our arguments we will first work with cosmologies involving purely expanding or contracting phases. Later we will discuss the cases where an expanding (contracting) phase can convert to a contracting (expanding) phase.
\section{The various cases in an ever expanding or ever contracting universe}
\label{excon}

In this section we will discuss about purely expanding or contracting phases in cosmology. Throughout the relevant interval, $t_i\le t \le t_f$, we assume that $\dot{a} \ne 0$. It is natural to assume that $a(t)>0$ inside the interval. The last condition may not be true for any time interval. For a contracting spacetime, the dynamics can produce $a(t_s)=0$ for some $t_i\le t_s \le t_f$. In such cases the system has reached a singularity. In this case one has to redefine the interval in such a way that the upper limit of the relevant interval, $t_f$, always remains smaller than $t_s$. In other words, our interval must be chosen in such a way that there are no singularities inside this interval. 
\subsection{Choosing the Friedmann equation and the continuity equation}
\label{fc}

When we choose the Friedmann equation and the continuity equation, we see that both of them are first order differential equations. They require two independent initial conditions: $a(t_i)$ and $\rho(t_i)$. We have not chosen the pressure equation and the acceleration equation in our scheme. Are those equations consistent with the dynamics produced from the two chosen equations? To answer that we proceed as follows.

To obtain the value of $\ddot{a}$ we first take a time derivative of both the sides of the Friedmann equation and get
\begin{eqnarray}
\frac{\dot{a}}{a}\left(\frac{2\ddot{a}}{a} - \frac{2\dot{a}^2}{a^2} - \frac{2k}{a^2}\right) 
= \frac{8\pi G}{3}\dot{\rho} = -8\pi G\frac{\dot{a}}{a}(\rho + P)\,,
\label{tdfr1}
\end{eqnarray}
where we have utilized the continuity equation to write the last step. Assuming $\dot{a}\ne 0$ in the interval $t_i \le t \le t_f$ we get
\begin{eqnarray}
\frac{2\ddot{a}}{a} - \frac{2\dot{a}^2}{a^2} - \frac{2k}{a^2} = -8\pi G(\rho + P)\,.
\label{tdfr2}
\end{eqnarray}
Writing the above equation as
\begin{eqnarray}
\frac{2\ddot{a}}{a} - \frac{2\dot{a}^2}{a^2} - \frac{2k}{a^2} + 8\pi G\rho= -8\pi G P\,,
\label{tdfr3}
\end{eqnarray}
and then utilizing the Friedmann equation, to replace $8\pi G\rho$, we get the pressure equation. This shows that the values of $a(t)$ and $\rho(t)$ generated from the Friedmann equation also satisfies the pressure equation for any $t$ inside the interval $[t_i\,,\,t_f]$ for cases where $\dot{a}\ne 0$ inside the time interval. One can again combine the pressure equation with the Friedmann equation to obtain the acceleration equation. As a consequence of this the generated values of $a(t)$ and $\rho(t)$ will also satisfy the acceleration equation. Hence all the relevant dynamical equations are satisfied simultaneously in this case. 
\subsection{Choosing the pressure equation and the continuity equation}
\label{pc}

In this scheme, we choose the pressure equation and the continuity equation as our basic dynamical equations. Here one of the equations is a second order differential equation, while the other one is a first order differential equation. Using the continuity equation we write the pressure equation as
\begin{eqnarray}
2\frac{\ddot{a}}{a} + \frac{\dot{a}^2}{a^2} + \frac{k}{a^2} = 8\pi G\left(\frac13
\dot{\rho}\frac{a}{\dot{a}} + \rho\right) + \Lambda\,,
\label{impfried}
\end{eqnarray}
where we have explicit appearance of $\dot{\rho}/\dot{a}$. The above equation always remain well defined as because $\dot{a}$ never vanishes in the time interval $[t_i\,,\,t_f]$.
Multiplying all the terms of the above equation by $\dot{a}/a$ we get
\begin{eqnarray}
2\frac{\ddot{a}\dot{a}}{a^2} + \frac{\dot{a}^3}{a^3} + \frac{k\dot{a}}{a^3} = \frac{8\pi G}{3}\dot{\rho} + 8\pi G \rho \frac{\dot{a}}{a} + \Lambda\frac{\dot{a}}{a}\,.
\label{friedpc}
\end{eqnarray}
The above equation can also be written as
$$\frac{d}{dt}\left(\frac{\dot{a}^2}{a^2}\right)+ \frac{3\dot{a}^3}{a^3} + \frac{k\dot{a}}{a^3} = \frac{8\pi G}{3}\dot{\rho} + 8\pi G \rho \frac{\dot{a}}{a} + \Lambda\frac{\dot{a}}{a}\,.$$
Subtracting
$$3\left(\frac{\dot{a}^3}{a^3} + \frac{k\dot{a}}{a^3}\right)$$
from both sides of the last equation, we get
\begin{eqnarray}
\frac{d}{dt}\left(\frac{\dot{a}^2}{a^2} + \frac{k}{a^2}\right)=\frac{8\pi G}{3}\frac{d\rho}{dt} + 3\frac{\dot{a}}{a}\left[\frac{8\pi G}{3}\rho + \frac{\Lambda}{3} -\left(\frac{\dot{a}^2}{a^2} + \frac{k}{a^2}\right)\right]\,.
\label{friedpc1}
\end{eqnarray}
At this stage, this equation does not convey anything interesting. It can become interesting if we define
\begin{eqnarray}
  F(t)\equiv \frac{8\pi G}{3}\rho + \frac{\Lambda}{3} -\left(\frac{\dot{a}^2}{a^2} + \frac{k}{a^2}\right)\,,
\label{fdef}  
\end{eqnarray}
and write Eq.~(\ref{friedpc1}) as
\begin{eqnarray}
\dot{F} + 3\frac{\dot{a}}{a}F=0\,.
\label{feqn}
\end{eqnarray}
This equation shows that the Friedmann equation may not hold when we obtain the dynamics of the system from the pressure equation and the continuity equation. If and only if $F(t)=0$ for all instants in the interval $[t_i\,,\,t_f]$, the Friedmann equation will be valid throughout the interval. In general for arbitrary initial conditions, $F(t)$ is never zero for all points inside the time interval. Consequently, for arbitrary initial conditions this dynamical evolution will not satisfy the Friedmann equation and the resultant dynamical evolution may not be a  valid solution of the cosmological evolution equations.
\subsection{Choosing the acceleration equation and the continuity equation}
\label{ac}

In this case we start with the acceleration equation and the continuity equation. This choice of equations is interesting as because none of the equations in our scheme have any dependence on $k$. Looking at the equations one cannot guess anything about the spatial curvature of the FLRW spacetime. Nonetheless, these equations are capable of producing dynamical evolution of the system when we impose reasonable initial conditions: $a(t_i)$, $\dot{a}(t_i)$ and $\rho(t_i)$. As because our dynamical system lacks any information about $k$, it may seem that it is impossible to obtain the other differential equations from these two equations as the other equations require specification of $k$. We will opine on this interesting question in the next subsection.

To proceed, we first rewrite the acceleration equation as
\begin{eqnarray}
\frac{\ddot{a}}{a}= -\frac{4\pi G}{3}(\rho + P) -\frac{8\pi G}{3}P + \frac{\Lambda}{3}\,.
\label{friedac}
\end{eqnarray}
Using the continuity equation, we can write the last equation as
\begin{eqnarray}
\frac{\ddot{a}}{a}= \frac{4\pi G}{9}\frac{a}{\dot{a}}\dot{\rho} -\frac{8\pi G}{3}P + \frac{\Lambda}{3}\,.
\label{friedac1}
\end{eqnarray}
This equation is also well defined throughout the interval $[t_i\,,\,t_f]$ as $\dot{a}$ never vanishes inside the interval. Multiplying both sides of the above equation by $2\dot{a}/a$ we get
\begin{eqnarray}
\frac{2\dot{a}\ddot{a}}{a^2}= \frac{8\pi G}{9}\dot{\rho} -\frac{16\pi G}{3}\frac{\dot{a}}{a}P + \frac{2\dot{a}\Lambda}{3a}\,.
\label{friedac2}
\end{eqnarray}
In the next step we will subtract
$$2\frac{\dot{a}}{a}\left(\frac{\dot{a}^2}{a^2} + \frac{\tilde{k}}{a^2}\right)\,,$$
from both sides of the above equation. At this level, $\tilde{k}$, is simply an arbitrary constant, whose value is not yet specified. Subtracting the aforementioned quantity from both sides of the last equation and manipulating some terms we get
\begin{eqnarray}
\frac{d}{dt}\left(\frac{\dot{a}^2}{a^2} + \frac{\tilde{k}}{a^2}\right)= \frac{8\pi G}{3}\dot{\rho} - \frac{16\pi G}{9}\dot{\rho}-\frac{16\pi G}{3}\frac{\dot{a}}{a}P + \frac{2\dot{a}\Lambda}{3a}
-2\frac{\dot{a}}{a}\left(\frac{\dot{a}^2}{a^2} + \frac{\tilde{k}}{a^2}\right)\,.
\label{friedac3}
\end{eqnarray}
The above equation can be written as
\begin{eqnarray}
\frac{d}{dt}\left(\frac{\dot{a}^2}{a^2} + \frac{\tilde{k}}{a^2} - \frac{8\pi G}{3}\rho - \frac{\Lambda}{3}\right)= -\frac{16\pi G}{9}\left(\dot{\rho}+\frac{3\dot{a}}{a}P\right) + \frac{2\dot{a}\Lambda}{3a}
-2\frac{\dot{a}}{a}\left(\frac{\dot{a}^2}{a^2} + \frac{\tilde{k}}{a^2}\right)\,.
\label{friedac4}
\end{eqnarray}
Defining $\tilde{F}(t)$ as
\begin{eqnarray}
\tilde{F}(t)\equiv \frac{8\pi G}{3}\rho + \frac{\Lambda}{3} -\left(\frac{\dot{a}^2}{a^2} + \frac{\tilde{k}}{a^2}\right)\,,
\label{fdefn}  
\end{eqnarray}
and using the continuity equation, we get
\begin{eqnarray}
\dot{\tilde{F}} &=& -\frac{16\pi G}{3}\frac{\dot{a}}{a}\rho - \frac{2\dot{a}\Lambda}{3a}
+2\frac{\dot{a}}{a}\left(\frac{\dot{a}^2}{a^2} + \frac{\tilde{k}}{a^2}\right)\nonumber\\
&=& -2\frac{\dot{a}}{a}\tilde{F}\,,
\label{friedac5}
\end{eqnarray}
which implies
\begin{eqnarray}
\dot{\tilde{F}} + 2\frac{\dot{a}}{a}\tilde{F}=0\,.
\label{feqnac}
\end{eqnarray}
This equation is similar to Eq.~(\ref{feqn}) but not exactly the same. The last equation is difficult to interpret as we still cannot fix the value of $\tilde{k}$ from our chosen set of equations. In such a case we cannot even guess whether the Friedmann equation is valid. 
\subsection{The link between all the choices is in the initial condition}

The pressure equation and continuity equation, as  well as, the acceleration equation and the continuity equation, requires three numbers in the initial condition: $a(t_i)$, $\dot{a}(t_i)$ and $\rho(t_i)$.
On the other hand the Friedmann equation and the continuity equation only requires two initial conditions: $a(t_i)$ and $\rho(t_i)$. 
If we treat all the differential equations as purely dynamical equations then it is really impossible to find out an equivalent initial condition.

The above problem can be tackled if we reinterpret the Friedmann equation as a constraint equation at the initial point $t_i$. The Friedmann equation is the only equation which  can be interpreted as an algebraic equation connecting the values of $\dot{a}(t_i)$, $a(t_i)$ and $\rho(t_i)$ for some values of $k$ and $\Lambda$.
Once we choose the initial conditions from the Friedmann equation itself, using it as a constraint equation, we see that the Friedmann equation remains valid throughout the time interval $[t_i\,,\,t_f]$. This happens because of the nature of Eq.~(\ref{feqn}). If the initial condition satisfies the Friedmann constraint then $F(t_i)=0$ and consequently $\dot{F}(t_i)=0$. As we chose the pressure equation and the continuity equation to generate the dynamics, we see that if the initial condition is chosen from the Friedmann constraint then the Friedmann equation remains valid throughout the time interval.

Next we focus our attention to the dynamics generated by the acceleration equation and the continuity equation.
If the initial condition is chosen from the Friedmann equation, with a specific value of $k$, then Eq.~(\ref{fdefn}) predicts $\tilde{k}=k$. The Friedmann constraint gives the information about the spatial curvature of the initial 3-dimensional hypersurface. Not only the initial condition fixes the value of $k$, now $\tilde{F}(t)=F(t)$, where $F(t)$ is defined as in Eq.~(\ref{fdef}). In this case also we see, from Eq.~(\ref{fdefn}), that if $\tilde{F}(t_i)=0$ then $\tilde{F}(t)=0$, at all instants in the interval $[t_i\,,\,t_f]$.  All the four relevant dynamical differential equations will have the same solution if we choose the initial condition via the Friedmann constraint.  

From the previous discussion it is seen that the Friedmann equation acts as a constraint equation, based on which the initial condition of cosmological dynamics is obtained. This constraint equation remains valid throughout the interval $[t_i\,,\,t_f]$ and consequently it can also be interpreted as a dynamical equation. The dual roles do not contradict because one can choose any time instant to be the initial time $t_i$. As a consequence the Friedmann equation becomes a dynamic constraint.   
\section{The cases where $\dot{a}(t_0)=0$ and $t_0$ is inside the interval $[t_i\,,\,t_f]$}
\label{dotaz}

The instant when $\dot{a}$ vanishes, is an instant when an expanding spacetime starts to contract or it can also be an instant, when an initially contracting spacetime starts to expand. The first case is called gravitational collapse and the second case is called a cosmological bounce, respectively. In these cases one has to interpret the results carefully. The first point to note is that both in Eq.~(\ref{impfried}) and in Eq.~(\ref{friedac1}) we have terms where $\dot{\rho}/\dot{a}$ appear explicitly. Apparently this terms may diverge when $\dot{a}$ vanishes. The fact is, none of these terms diverge at any moment inside the relevant time interval. At the instant when $\dot{a}$ vanishes, $\dot{\rho}$ also vanishes and  their ratio remains finite. From the continuity equation we get
$$\frac{\dot{\rho}}{\dot{a}}=-\frac{3(\rho + P)}{a}\,.$$
Even when both $\dot{\rho}$ and $\dot{a}$ tends to zero, inside the interval $[t_i\,,\,t_f]$, their ratio remains finite. 
\subsection{When $t_0=t_i$}

In this case we have $\dot{a}(t_i)=0$ at the initial point. That a solution exists in the present case can be shown in the following way. We know that the Friedmann equation and the continuity equation can generate a continuous $a(t)$ and $\rho(t)$ in the time interval  $[t_i\,,\,t_f]$. Suppose instead of $t_i$, we start at $t_i+\epsilon$, where $\epsilon$ is an arbitrarily small positive constant with dimension of time. We assume that $\dot{a}(t_i)=0$ and consequently $\dot{a}$ never vanishes inside the interval $[t_i+\epsilon\,,\,t_f]$. The initial conditions at $t=t_i+\epsilon$  are now generated from the known solution coming out from the Friedmann equation and continuity equation. All the generated solutions for different (but arbitrarily small) $\epsilon$ (possibly using other pairs of differential equations) smoothly matches with the solution defined in the full time interval for regions $t_i+\epsilon \le t \le t_f$. One can make $\epsilon$ arbitrarily small and continuity of our generated solution implies that in the limit $\epsilon \to 0$ we will also have a unique solution of the Einstein equation. The solutions from the different pairs coincide at $t_i=t_0$ as the pressure equation and the continuity equation starts with initial conditions following the Friedmann constraint. 

For the other cases related to Eq.~(\ref{feqn}) and Eq.~(\ref{feqnac}), we see that if $\dot{a}(t_i)=0$, it seems that both these equations can be satisfied for any $F(t_i)$ as long as $\dot{F}(t_i)=0$. But this arbitrariness disappears once we use the Friedmann constraint initially and fix $F(t_i)=0$. As a result of this the Friedmann equation will be valid for all points inside the relevant time interval (if we start from pairs of equations not containing the Friedmann equation) and we will always get a definite solution to the Einstein equation.
\subsection{When $t_i < t_0 < t_f$}

Next we discuss the case where $\dot{a}(t_0)=0$ where $t_0$ is an intermediate point in the relevant time interval. In this case we can obtain a solution to the Einstein equation if our initial condition satisfies the Friedmann constraint. We can follow a solution of the Einstein's equation involving gravitational expansion and a subsequent gravitational contraction in the interval $[t_i\,,\,t_f]$ by using the Friedmann equation and the continuity equation. The turn-around happens at $t=t_0$ when $\dot{a}(t_0)=0$. Is this solution also obtained by the other pairs of equations? Before we proceed we observe that all the relevant dynamical equations, namely the Friedmann equation, pressure equation, acceleration equation and the continuity equation are symmetric around $t=0$, by which we mean
\begin{eqnarray}
a(t)=a(-t)\,,\,\,\,\,\rho(t)=\rho(-t)\,,\,\,\,\,\dot{a}(t)=-\dot{a}(-t)\,,\,\,\,\,\ddot{a}(t)=\ddot{a}(-t)\,,\,\,(t \ge 0)\,,
\label{trev}
\end{eqnarray}
when $\dot{a}(t=0)=0$. We will use these properties and shift the origin of time to $t_0$ so that what  we previously called the time instant $t_0$ becomes the origin of time $t=0$.  The initial time interval now becomes $[t_i-t_0\,,\,t_f-t_0]$. In shifted time we have $\dot{a}(0)=0$ which implies a symmetric gravitational collapse. The initial conditions can be well specified in the shifted time as $\rho(0)$ and $a(0)$. Now using any pair of differential equations we can generate the solution of the Einstein equation in the expanding phase, in the interval $[t_i-t_0,0]$. In this interval all the solutions will agree if all of them satisfy the Friedmann constraint initially. The symmetry of the Einstein equations now guarantee that these solutions must also agree on the contracting phase. If one now wants to go back to the old time interval, one has to just shift the time origin back to $t_i$. 

\section{On the logical consistency of the theory}

The results of this article are related to the initial value problem and its solution in cosmology. A formal discussion on the initial value problem in cosmology, using techniques of differential geometry, can be found in Refs.~\cite{Poisson:2009pwt, Wald:1984rg}. In this article we focus on cosmological dynamics to infer upon the initial value problem in cosmology. In the rest of this section we discuss about the logical consistency of cosmological dynamics in GR.

If we leave out quantum effects, we may safely assume that
there cannot be matter creation from the gravitational field. In such
a case we must have some condition which predicts $\rho(t)=0$ in the interval $[t_i\,,\,t_f]$ if $\rho(t_i)=0$. Such a relation is there in GR, as the theory is formulated in such a way that the desired condition is produced due to a tensor identity, called the (second) Bianchi identity. This identity predicts the conservation of matter energy-momentum tensor, which yields a first order differential equation: where $\dot{\rho}(t)\propto \rho(t)$.
This is a very general requirement, as for such a relation if $\rho$ is zero initially then $\dot{\rho}$ is also zero initially and hence the matter energy
density must remain zero for all times.

As the Einstein equations (Friedmann and pressure equations) in cosmology do not contain a time derivative of the energy density of the fluid
so those equations does not naturally generate a differential equation for $\rho(t)$, but we know from general assumptions that such an equation (as predicted by the Bianchi identity) is inevitable. In such a case there is only one way the first order differential equation for $\rho(t)$ can appear, that intended equation must be derivable from the available Einstein equations.
This implies that one of the Einstein equations, containing $\rho$ on the right hand side, must have to be a first order differential equation in the scale-factor, $a(t)$. It turns out that equation is the Friedmann equation.
The Friedmann equation cannot be a second order differential equation, for $a(t)$, as in that case a differentiation of this equation will produce a third order derivative of the scale-factor in the resulting equation which cannot be cancelled by any terms in the other Einstein equation for pressure.
As because GR, as a well defined field theory, contains second order derivative terms of the scale-factor, the other equation for pressure must have to be a second order differential equation for $a(t)$. 

These arguments show the subtle relation between the redundancy problem and the special role of the Bianchi identity in GR. The presence of the Bianchi identity compels the FLRW spacetime to have different orders of differential equations which leads to the Friedmann constraint for fixing equivalent initial conditions.  

\section{Conclusion}

In this paper we have analyzed the dynamical structure of cosmological evolution in GR. In particular we addressed the issues related to the redundancy of cosmological evolution equations and the relationship of this redundancy with the initial value problem in cosmology.  There is a branch of cosmology, called Newtonian cosmology, where one obtains the basic cosmological equations in GR in a dust dominated universe primarily using some ad-hoc assumptions on the nature of spacetime and semi-Newtonian conceptions \cite{milne, peebles, tipler}. As the basic equations of cosmology in these models remain the same as in GR, one faces the same redundancy problem in these cases and our solution of the redundancy problem works in Newtonian cosmology.

It is seen that by focusing our attention on the solvability of the dynamical system we can learn important properties of the Einstein equation in cosmology and unravel the special role of the Bianchi identity. Throughout we have applied a bottom-up approach and it seems that this approach yields interesting results. The strength of our approach is related to the simplicity and accessibility of the analysis.



\begin{thebibliography}{100}

\bibitem{Kolb:1990vq}
E.~W.~Kolb and M.~S.~Turner,
Front. Phys. \textbf{69}, 1-547 (1990)
doi:10.1201/9780429492860

\bibitem{Mukhanov:2005sc}
V.~Mukhanov,
``Physical Foundations of Cosmology,''
Cambridge University Press, 2005,
ISBN 978-0-521-56398-7

\bibitem{Dodelson:2003ft}
S.~Dodelson,
``Modern Cosmology,''
Academic Press, 2003,
ISBN 978-0-12-219141-1

\bibitem{Zee:2013dea}
A.~Zee,
``Einstein Gravity in a Nutshell,''
Princeton University Press, 2013,
ISBN 978-0-691-14558-7

\bibitem{Hartle:2021pel}
J.~B.~Hartle,
``Gravity,''
Cambridge University Press, 2021,
ISBN 978-1-00-904260-4

\bibitem{Weinberg:1972kfs}
S.~Weinberg,
``Gravitation and Cosmology: Principles and Applications of the General Theory of Relativity,''
John Wiley and Sons, 1972,
ISBN 978-0-471-92567-5, 978-0-471-92567-5

\bibitem{Nemiroff:2007xs}
R.~J.~Nemiroff and B.~Patla,
``Adventures in Friedmann Cosmology: An Educationally Detailed Expansion of the Cosmological Friedmann Equations,''
Am. J. Phys. \textbf{76}, 265-276 (2008)
doi:10.1119/1.2830536 

\bibitem{Sonego:2011rb}
S.~Sonego and V.~Talamini,
``Qualitative study of perfect-fluid Friedmann-Lemaitre-Robertson-Walker models with a cosmological constant,''
Am. J. Phys. \textbf{80}, 670-679 (2012)
doi:10.1119/1.4731258

\bibitem{Faraoni:1999qu}
V.~Faraoni,
``Solving for the dynamics of the universe,''
Am. J. Phys. \textbf{67}, 732 (1999)
doi:10.1119/1.19361

\bibitem{ReFiorentin:2023bhk}
M.~Re Fiorentin and S.~Re Fiorentin,
``Cosmological horizons,''
Am. J. Phys. \textbf{91}, no.8, 644 (2023)
doi:10.1119/5.0127840

\bibitem{Christiansen:2012fy}
J.~Christiansen and A.~Siver,
``Computing Accurate Age and Distance Factors in Cosmology,''
Am. J. Phys. \textbf{80}, 367 (2012)
doi:10.1119/1.3698352

\bibitem{Poisson:2009pwt}
E.~Poisson,
``A Relativist's Toolkit: The Mathematics of Black-Hole Mechanics,''
Cambridge University Press, 2009,
doi:10.1017/CBO9780511606601


\bibitem{Wald:1984rg}
R.~M.~Wald,
``General Relativity,''
Chicago Univ. Pr., 1984,
doi:10.7208/chicago/9780226870373.001.0001








\bibitem{milne}
W.~H.~McCrea and E.~Milne,
``Newtonian Universes and the Curvature of Space,''
Q. J. Maths. \textbf{5}, 73-80 (1934)


\bibitem{peebles}
C.~Callan, R.~H.~Dicke and P.~J.~E.~Peebles,
``Cosmology and Newtonian Mechanics,''
Am. J. Phys. \textbf{33}, 105 (1965)
doi:10.1119/1.1971256


\bibitem{tipler}
Frank~J.~Tipler,
``Rigorous Newtonian Cosmology,''
Am. J. Phys. \textbf{64}, 1311-1315 (1996)
doi:10.1119/1.18398

\end{thebibliography}
\end{document}